\PassOptionsToPackage{unicode}{hyperref}
\PassOptionsToPackage{hyphens}{url}
\PassOptionsToPackage{dvipsnames,svgnames,x11names}{xcolor}
\documentclass[
]{article}
\usepackage{amsmath,amssymb}
\usepackage{lmodern}
\usepackage{iftex}
\ifPDFTeX
  \usepackage[T1]{fontenc}
  \usepackage[utf8]{inputenc}
  \usepackage{textcomp} 
\else 
  \usepackage{unicode-math}
  \defaultfontfeatures{Scale=MatchLowercase}
  \defaultfontfeatures[\rmfamily]{Ligatures=TeX,Scale=1}
\fi
\IfFileExists{upquote.sty}{\usepackage{upquote}}{}
\IfFileExists{microtype.sty}{
  \usepackage[]{microtype}
  \UseMicrotypeSet[protrusion]{basicmath} 
}{}
\makeatletter
\@ifundefined{KOMAClassName}{
  \IfFileExists{parskip.sty}{%
    \usepackage{parskip}
  }{
    \setlength{\parindent}{0pt}
    \setlength{\parskip}{6pt plus 2pt minus 1pt}}
}{
  \KOMAoptions{parskip=half}}
\makeatother
\usepackage{xcolor}
\NewDocumentCommand\citeproctext{}{}
\NewDocumentCommand\citeproc{mm}{%
  \begingroup\def\citeproctext{#2}\cite{#1}\endgroup}
\makeatletter
 \let\@cite@ofmt\@firstofone
 \def\@biblabel#1{}
 \def\@cite#1#2{{#1\if@tempswa , #2\fi}}
\makeatother
\newlength{\cslhangindent}
\newlength{\csllabelwidth}
\newenvironment{CSLReferences}[2] 
 {\begin{list}{}{%
  \setlength{\itemindent}{0pt}
  \setlength{\leftmargin}{0pt}
  \setlength{\parsep}{0pt}
  \ifodd #1
   \setlength{\leftmargin}{\cslhangindent}
   \setlength{\itemindent}{-1\cslhangindent}
  \fi
  \setlength{\itemsep}{#2\baselineskip}}}
 {\end{list}}
\usepackage{calc}

\ifLuaTeX
\usepackage[bidi=basic]{babel}
\else
\usepackage[bidi=default]{babel}
\fi
\babelprovide[main,import]{american}

\def\languageshorthands#1{}
\ifLuaTeX
  \usepackage{selnolig}  
\fi
\IfFileExists{bookmark.sty}{\usepackage{bookmark}}{\usepackage{hyperref}}
\IfFileExists{xurl.sty}{\usepackage{xurl}}{} 
\hypersetup{
  pdftitle={TopologicalNumbers.jl: A Julia package for topological
number computation},
  pdfauthor={Keisuke Adachi, Minoru Kanega},
  pdflang={en-US},
  colorlinks=true,
  linkcolor={Maroon},
  filecolor={Maroon},
  citecolor={Blue},
  urlcolor={Blue},
  pdfcreator={LaTeX via pandoc}}

\title{TopologicalNumbers.jl: A Julia package for topological number
computation}

\definecolor{c53baa1}{RGB}{83,186,161}
\definecolor{c202826}{RGB}{32,40,38}

\usepackage[affil-it]{authblk}
\usepackage{orcidlink}
\author[1,2%
  *%
  \ensuremath\mathparagraph]{Keisuke Adachi%
    \,\orcidlink{0009-0004-0195-7952}\,%
    }
\author[2%
  *%
  \ensuremath\mathparagraph]{Minoru Kanega%
    \,\orcidlink{0009-0008-4623-8010}\,%
    }

\affil[1]{Department of Physics, Ibaraki University, Mito, Ibaraki,
Japan%
  }
\affil[2]{Department of Physics, Chiba University, Chiba, Japan%
  }
\affil[$\mathparagraph$]{Corresponding author: %
}
\affil[*]{These authors contributed equally.}
\date{}

\begin{document}
\maketitle

\section{Summary}\label{summary}

\texttt{TopologicalNumbers.jl} is an open-source Julia package designed
to calculate topological invariants --- mathematical quantities that
characterize the properties of materials in condensed matter physics.
These invariants, such as the Chern number and the \(\mathbb{Z}_2\)
invariant, are crucial for understanding exotic materials like
topological insulators and superconductors, which have potential
applications in advanced electronics, spintronics, and quantum computing
(\citeproc{ref-Hasan2010Colloquium}{Hasan \& Kane, 2010};
\citeproc{ref-Nayak2008NonAbelian}{Nayak et al., 2008};
\citeproc{ref-Qi2011Topological}{Qi \& Zhang, 2011}). This package
provides researchers and educators with an easy-to-use and efficient
toolset to compute these invariants across various dimensions and
symmetry classes, facilitating the exploration and discovery of new
topological phases of matter.

\section{Statement of need}\label{statement-of-need}

Understanding the properties of materials is essential in solid-state
physics. For example, electrical conductivity is a key physical quantity
indicating how well a material conducts electric current. Typically,
when a weak electric field is applied to a material, if quantum
eigenstates exist in the bulk into which electrons can transition, the
material exhibits finite electrical conductivity and behaves as a metal.
Conversely, if such states do not exist, the electrical conductivity is
low, and the material behaves as an insulator. Since the 1980s, it has
been revealed that certain materials exhibit states in which the bulk is
insulating while the material's surface has conducting electronic
states~(\citeproc{ref-Hasan2010Colloquium}{Hasan \& Kane, 2010};
\citeproc{ref-Qi2011Topological}{Qi \& Zhang, 2011}). These materials
are known as topological electronic systems, including quantum Hall
insulators and topological insulators. Due to these novel properties,
extensive research has been conducted to identify candidate materials
and evaluate their characteristics.

The features of surface conducting states are determined by the topology
of quantum eigenstates in momentum space. Topological numbers, such as
the first Chern number, the second Chern number, and the
\(\mathbb{Z}_2\) invariant, are used to characterize these
properties~(\citeproc{ref-Kane2005Z_2}{Kane \& Mele, 2005};
\citeproc{ref-Thouless1982Quantized}{Thouless et al., 1982}). A typical
example is the quantum Hall effect, where applying a weak electric field
to a two-dimensional material results in a quantized finite electrical
conductivity (Hall conductivity) perpendicular to the applied
field~(\citeproc{ref-Thouless1982Quantized}{Thouless et al., 1982}). The
Hall conductivity \(\sigma_{xy}\) is characterized by the first Chern
number \(\nu \in \mathbb{Z}\) and is given by
\(\sigma_{xy} = \frac{e^{2}}{h} \nu\), where \(e\) is the elementary
charge and \(h\) is Planck's constant. Other topological numbers
similarly serve as important physical quantities that characterize
systems, depending on their dimensions and symmetry
classes~(\citeproc{ref-Ryu2010Topological}{Ryu et al., 2010}).

Obtaining topological numbers often requires extensive numerical
calculations, which may demand considerable computational effort before
achieving convergence. Therefore, creating tools that simplify the
computation of these quantities will advance research on topological
phases of matter. Several methods have been developed to enable
efficient computation of certain topological numbers
(\citeproc{ref-Fukui2005Chern}{Fukui et al., 2005};
\citeproc{ref-Fukui2007Quantum}{Fukui \& Hatsugai, 2007};
\citeproc{ref-Mochol-Grzelak2018Efficient}{Mochol-Grzelak et al., 2018};
\citeproc{ref-Shiozaki2023discrete}{Shiozaki, 2023}). However, since
each method is typically specialized for specific dimensions or symmetry
classes, one must often implement algorithms separately for each
problem. Our project, \texttt{TopologicalNumbers.jl}, aims to provide a
package that can efficiently and easily compute topological numbers
across various dimensions and symmetry classes.

This package currently includes several methods for calculating
topological numbers. The first is the Fukui--Hatsugai--Suzuki (FHS)
method for computing the first Chern number in two-dimensional
solid-state systems (\citeproc{ref-Fukui2005Chern}{Fukui et al., 2005}).
The first Chern number is obtained by integrating the Berry curvature,
derived from the Hamiltonian's eigenstates, over the Brillouin zone. The
FHS method enables efficient computation by discretizing the Berry
curvature in the Brillouin zone. Several methods based on the FHS
approach have been proposed to compute various topological numbers. One
such method calculates the second Chern number in four-dimensional
systems (\citeproc{ref-Mochol-Grzelak2018Efficient}{Mochol-Grzelak et
al., 2018}). The \(\mathbb{Z}_2\) invariant can be computed in
two-dimensional systems with time-reversal symmetry
(\citeproc{ref-Fukui2007Quantum}{Fukui \& Hatsugai, 2007};
\citeproc{ref-Shiozaki2023discrete}{Shiozaki, 2023}). The FHS method
also applies to identifying Weyl points and Weyl nodes in
three-dimensional systems (\citeproc{ref-Du2017Emergence}{Du et al.,
2017}; \citeproc{ref-Hirayama2015Weyl}{Hirayama et al., 2015},
\citeproc{ref-Hirayama2018Topological}{2018};
\citeproc{ref-Yang2011Quantum}{Yang et al., 2011}).

Currently, there is no comprehensive Julia package that implements all
these calculation methods. On other platforms, software packages using
different approaches --- such as those based on Wannier charge
centers~(\citeproc{ref-Soluyanov2011Computing}{Soluyanov, 2011}) or
Wilson loops~(\citeproc{ref-Yu2011Equivalent}{Yu et al., 2011}) --- are
available. For example, \texttt{Z2Pack}
(\citeproc{ref-Gresch2017Z2Pack}{Gresch et al., 2017}) is a Python-based
tool widely used for calculating the \(\mathbb{Z}_2\) invariant and the
first Chern number. \texttt{WannierTools}
(\citeproc{ref-Wu2018WannierTools}{Wu et al., 2018}) offers powerful
features for analyzing topological materials but is implemented in
Fortran, which may pose a steep learning curve for some users.

\texttt{TopologicalNumbers.jl} distinguishes itself by providing an
efficient, pure Julia implementation. Julia is known for its high
performance and user-friendly syntax. This package supports various
topological invariants across multiple dimensions and symmetry classes,
including the first and second Chern numbers and the \(\mathbb{Z}_2\)
invariant. It also offers parallel computing capabilities through
\texttt{MPI.jl}, enhancing computational efficiency for large-scale
problems. By leveraging Julia's multiple dispatch feature, we adopt a
consistent interface using the \texttt{Problem}, \texttt{Algorithm}, and
\texttt{solve} style --- similar to
\texttt{DifferentialEquations.jl}~(\citeproc{ref-Rackauckas2017DifferentialEquationsjl}{Rackauckas
\& Nie, 2017}) --- to improve extensibility. With these features,
\texttt{TopologicalNumbers.jl} achieves a unique balance of performance,
usability, maintainability, and extensibility, providing an alternative
perspective rather than directly competing with other libraries.

Additionally, to compute the \(\mathbb{Z}_2\) invariant, which requires
calculating the Pfaffian, we have ported \texttt{PFAPACK} to Julia.
\texttt{PFAPACK} is a Fortran/C++/Python library for computing the
Pfaffian of skew-symmetric matrices
(\citeproc{ref-Wimmer2012Algorithm}{Wimmer, 2012}). Our package includes
pure Julia implementations of all originally provided functions. While
\texttt{SkewLinearAlgebra.jl} exists as an official Julia package for
computing the Pfaffian of real skew-symmetric matrices, to our
knowledge, \texttt{TopologicalNumbers.jl} is the first official package
to offer a pure Julia implementation that handles complex skew-symmetric
matrices.

\section{Usage}\label{usage}

Users can easily compute topological numbers using the various methods
included in this package. In the simplest case, they need only provide a
function that returns the Hamiltonian matrix as a function of the wave
numbers. Computations are performed by creating the corresponding
\texttt{Problem} instance and calling the \texttt{solve} function
(\texttt{solve(Problem)}). The package also provides the
\texttt{calcPhaseDiagram} function, which enables the computation of
topological numbers in one-dimensional or two-dimensional parameter
spaces by specifying the \texttt{Problem} and parameter ranges
(\texttt{calcPhaseDiagram(Problem,\ range...)}).

Moreover, utility functions such as \texttt{showBand}, \texttt{plot1D},
and \texttt{plot2D} are available for visualizing energy band structures
and phase diagrams. We also offer various model Hamiltonians, such as
the Su--Schrieffer--Heeger (SSH) model (\citeproc{ref-Su1979Solitons}{Su
et al., 1979}) and the Haldane model
(\citeproc{ref-Haldane1988Model}{Haldane, 1988}), allowing users to
quickly test functionalities and learn how to use these features.

\section{Acknowledgements}\label{acknowledgements}

The authors are grateful to Takahiro Fukui for fruitful discussions. M.
K. was supported by JST, the establishment of university fellowships
towards the creation of science technology innovation (Grant
No.~JPMJFS2107), and by JST SPRING (Grant No.~JPMJSP2109).

\section*{References}\label{references}
\addcontentsline{toc}{section}{References}

\phantomsection\label{refs}


\begin{CSLReferences}{1}{0}
\bibitem[\citeproctext]{ref-Du2017Emergence}
Du, Y., Bo, X., Wang, D., Kan, E., Duan, C.-G., Savrasov, S. Y., \& Wan,
X. (2017). Emergence of topological nodal lines and type-{II Weyl} nodes
in the strong spin-orbit coupling system InNbX2 (x = s,se). \emph{Phys.
Rev. B}, \emph{96}(23), 235152.
\url{https://doi.org/10.1103/PhysRevB.96.235152}

\bibitem[\citeproctext]{ref-Fukui2007Quantum}
Fukui, T., \& Hatsugai, Y. (2007). Quantum {Spin Hall Effect} in {Three
Dimensional Materials}: {Lattice Computation} of {Z2 Topological
Invariants} and {Its Application} to {Bi} and {Sb}. \emph{J. Phys. Soc.
Jpn.}, \emph{76}(5), 053702.
\url{https://doi.org/10.1143/JPSJ.76.053702}

\bibitem[\citeproctext]{ref-Fukui2005Chern}
Fukui, T., Hatsugai, Y., \& Suzuki, H. (2005). Chern {Numbers} in
{Discretized Brillouin Zone}: {Efficient Method} of {Computing} ({Spin})
{Hall Conductances}. \emph{J. Phys. Soc. Jpn.}, \emph{74}(6),
1674--1677. \url{https://doi.org/10.1143/JPSJ.74.1674}

\bibitem[\citeproctext]{ref-Gresch2017Z2Pack}
Gresch, D., Autès, G., Yazyev, O. V., Troyer, M., Vanderbilt, D.,
Bernevig, B. A., \& Soluyanov, A. A. (2017). {Z2Pack}: {Numerical}
implementation of hybrid {Wannier} centers for identifying topological
materials. \emph{Phys. Rev. B}, \emph{95}(7), 075146.
\url{https://doi.org/10.1103/PhysRevB.95.075146}

\bibitem[\citeproctext]{ref-Haldane1988Model}
Haldane, F. D. M. (1988). Model for a {Quantum Hall Effect} without
{Landau Levels}: {Condensed-Matter Realization} of the "{Parity
Anomaly}". \emph{Phys. Rev. Lett.}, \emph{61}(18), 2015--2018.
\url{https://doi.org/10.1103/PhysRevLett.61.2015}

\bibitem[\citeproctext]{ref-Hasan2010Colloquium}
Hasan, M. Z., \& Kane, C. L. (2010). Colloquium: {Topological}
insulators. \emph{Rev. Mod. Phys.}, \emph{82}(4), 3045--3067.
\url{https://doi.org/10.1103/RevModPhys.82.3045}

\bibitem[\citeproctext]{ref-Hirayama2015Weyl}
Hirayama, M., Okugawa, R., Ishibashi, S., Murakami, S., \& Miyake, T.
(2015). Weyl {Node} and {Spin Texture} in {Trigonal Tellurium} and
{Selenium}. \emph{Phys. Rev. Lett.}, \emph{114}(20), 206401.
\url{https://doi.org/10.1103/PhysRevLett.114.206401}

\bibitem[\citeproctext]{ref-Hirayama2018Topological}
Hirayama, M., Okugawa, R., \& Murakami, S. (2018). Topological
{Semimetals Studied} by {Ab Initio Calculations}. \emph{J. Phys. Soc.
Jpn.}, \emph{87}(4), 041002.
\url{https://doi.org/10.7566/JPSJ.87.041002}

\bibitem[\citeproctext]{ref-Kane2005Z_2}
Kane, C. L., \& Mele, E. J. (2005). \(\mathbb{Z}_2\) {Topological Order}
and the {Quantum Spin Hall Effect}. \emph{Phys. Rev. Lett.},
\emph{95}(14), 146802.
\url{https://doi.org/10.1103/PhysRevLett.95.146802}

\bibitem[\citeproctext]{ref-Mochol-Grzelak2018Efficient}
Mochol-Grzelak, M., Dauphin, A., Celi, A., \& Lewenstein, M. (2018).
Efficient algorithm to compute the second {Chern} number in four
dimensional systems. \emph{Quantum Sci. Technol.}, \emph{4}(1), 014009.
\url{https://doi.org/10.1088/2058-9565/aae93b}

\bibitem[\citeproctext]{ref-Nayak2008NonAbelian}
Nayak, C., Simon, S. H., Stern, A., Freedman, M., \& Das Sarma, S.
(2008). Non-{Abelian} anyons and topological quantum computation.
\emph{Rev. Mod. Phys.}, \emph{80}(3), 1083--1159.
\url{https://doi.org/10.1103/RevModPhys.80.1083}

\bibitem[\citeproctext]{ref-Qi2011Topological}
Qi, X.-L., \& Zhang, S.-C. (2011). Topological insulators and
superconductors. \emph{Rev. Mod. Phys.}, \emph{83}(4), 1057--1110.
\url{https://doi.org/10.1103/RevModPhys.83.1057}

\bibitem[\citeproctext]{ref-Rackauckas2017DifferentialEquationsjl}
Rackauckas, C., \& Nie, Q. (2017). {DifferentialEquations}.jl -- {A
Performant} and {Feature-Rich Ecosystem} for {Solving Differential
Equations} in {Julia}. \emph{J. Open Res. Softw.}, \emph{5}(1), 15--15.
\url{https://doi.org/10.5334/jors.151}

\bibitem[\citeproctext]{ref-Ryu2010Topological}
Ryu, S., Schnyder, A. P., Furusaki, A., \& Ludwig, A. W. W. (2010).
Topological insulators and superconductors: Tenfold way and dimensional
hierarchy. \emph{New J. Phys.}, \emph{12}(6), 065010.
\url{https://doi.org/10.1088/1367-2630/12/6/065010}

\bibitem[\citeproctext]{ref-Shiozaki2023discrete}
Shiozaki, K. (2023). \emph{A discrete formulation of the {Kane-Mele}
\({\mathbb Z}_2\) invariant} (No. arXiv:2305.05615). {arXiv}.
\url{https://doi.org/10.48550/arXiv.2305.05615}

\bibitem[\citeproctext]{ref-Soluyanov2011Computing}
Soluyanov, A. A. (2011). Computing topological invariants without
inversion symmetry. \emph{Phys. Rev. B}, \emph{83}(23).
\url{https://doi.org/10.1103/PhysRevB.83.235401}

\bibitem[\citeproctext]{ref-Su1979Solitons}
Su, W. P., Schrieffer, J. R., \& Heeger, A. J. (1979). Solitons in
{Polyacetylene}. \emph{Phys. Rev. Lett.}, \emph{42}(25), 1698--1701.
\url{https://doi.org/10.1103/PhysRevLett.42.1698}

\bibitem[\citeproctext]{ref-Thouless1982Quantized}
Thouless, D. J., Kohmoto, M., Nightingale, M. P., \& den Nijs, M.
(1982). Quantized {Hall Conductance} in a {Two-Dimensional Periodic
Potential}. \emph{Phys. Rev. Lett.}, \emph{49}(6), 405--408.
\url{https://doi.org/10.1103/PhysRevLett.49.405}

\bibitem[\citeproctext]{ref-Wimmer2012Algorithm}
Wimmer, M. (2012). Algorithm 923: {Efficient Numerical Computation} of
the {Pfaffian} for {Dense} and {Banded Skew-Symmetric Matrices}.
\emph{ACM Trans. Math. Softw.}, \emph{38}(4), 30:1--30:17.
\url{https://doi.org/10.1145/2331130.2331138}

\bibitem[\citeproctext]{ref-Wu2018WannierTools}
Wu, Q., Zhang, S., Song, H.-F., Troyer, M., \& Soluyanov, A. A. (2018).
{WannierTools}: {An} open-source software package for novel topological
materials. \emph{Comput. Phys. Commun.}, \emph{224}, 405--416.
\url{https://doi.org/10.1016/j.cpc.2017.09.033}

\bibitem[\citeproctext]{ref-Yang2011Quantum}
Yang, K.-Y., Lu, Y.-M., \& Ran, Y. (2011). Quantum {Hall} effects in a
{Weyl} semimetal: {Possible} application in pyrochlore iridates.
\emph{Phys. Rev. B}, \emph{84}(7), 075129.
\url{https://doi.org/10.1103/PhysRevB.84.075129}

\bibitem[\citeproctext]{ref-Yu2011Equivalent}
Yu, R., Qi, X. L., Bernevig, A., Fang, Z., \& Dai, X. (2011). Equivalent
expression of \(\mathbb{Z}_2\) topological invariant for band insulators
using the non-{Abelian Berry} connection. \emph{Phys. Rev. B},
\emph{84}(7), 075119. \url{https://doi.org/10.1103/PhysRevB.84.075119}

\end{CSLReferences}
\end{document}